\documentclass[aps, prl, twocolumn, superscriptaddress, floatfix, showpacs,longbibliography]{revtex4-1}

\usepackage{latexsym}
\usepackage{bm}
\usepackage{graphicx}        
\usepackage{amssymb}
\usepackage{amsmath}
\usepackage[usenames,dvipsnames]{xcolor}
\usepackage[colorlinks=true, urlcolor=urlcolor, linkcolor=linkcolor, bookmarks=true, citecolor=blue, breaklinks=true]{hyperref}

\usepackage{subfigure}

\colorlet{notecolor2}{olive}
\definecolor{notecolor}{RGB}{150,20,40}
\definecolor{linkcolor}{RGB}{20,90,15}
\definecolor{urlcolor}{RGB}{150,20,40}

\begin{document}

\title{Heteronuclear Limit of Strong-Field Ionization: Fragmentation of HeH$^+$ by Intense Ultrashort Laser Pulses}

\author{Philipp\,Wustelt}
\email{philipp.wustelt@uni-jena.de}
\affiliation{Institute of Optics and Quantum Electronics, Friedrich Schiller University Jena, 07743 Jena, Germany}
\affiliation{Helmholtz Institute Jena, Fr\"obelstieg 3, 07743 Jena, Germany}

\author{Florian\,Oppermann}
\affiliation{Institut f\"ur Theoretische Physik,  Leibniz Universit\"at  Hannover, Appelstra\ss e 2, 30167 Hannover, Germany}

\author{Lun\,Yue}
\affiliation{Institute of Physical Chemistry, Friedrich Schiller
University Jena, Helmholtzweg 4, 07743 Jena, Germany}

\author{Max\,M\"oller}
\affiliation{Institute of Optics and Quantum Electronics, Friedrich Schiller University Jena, 07743 Jena, Germany}
\affiliation{Helmholtz Institute Jena, Fr\"obelstieg 3, 07743 Jena, Germany}

\author{Thomas\,St\"ohlker}
\affiliation{Institute of Optics and Quantum Electronics, Friedrich Schiller University Jena, 07743 Jena, Germany}
\affiliation{Helmholtz Institute Jena, Fr\"obelstieg 3, 07743 Jena, Germany}

\author{Manfred\,Lein}
\affiliation{Institut f\"ur Theoretische Physik,  Leibniz Universit\"at  Hannover, Appelstra\ss e 2, 30167 Hannover, Germany}

\author{Stefanie\,Gr\"afe}
\affiliation{Institute of Physical Chemistry, Friedrich Schiller
University Jena, Helmholtzweg 4, 07743 Jena, Germany}

\author{Gerhard\,G.\,Paulus}
\email{gerhard.paulus@uni-jena.de}
\affiliation{Institute of Optics and Quantum Electronics, Friedrich Schiller University Jena, 07743 Jena, Germany}
\affiliation{Helmholtz Institute Jena, Fr\"obelstieg 3, 07743 Jena, Germany}

\author{A.\,Max\,Sayler}
\affiliation{Institute of Optics and Quantum Electronics, Friedrich Schiller University Jena, 07743 Jena, Germany}
\affiliation{Helmholtz Institute Jena, Fr\"obelstieg 3, 07743 Jena, Germany}

\date{\today}

\begin{abstract}
The laser-induced fragmentation dynamics of this most fundamental polar molecule HeH$^+$ are measured using an ion beam of helium hydride and an isotopologue at various wavelengths and intensities. In contrast to the prevailing interpretation of strong-field fragmentation, in which stretching of the molecule results primarily from laser-induced electronic excitation, experiment and theory for nonionizing dissociation, single ionization and double ionization both show that the direct vibrational excitation plays the decisive role here.  We are able to reconstruct fragmentation pathways and determine the times at which each ionization step occurs as well as the bond length evolution before the electron removal. The dynamics of this extremely asymmetric molecule contrast the well-known symmetric systems leading to a more general picture of strong-field molecular dynamics and facilitating interpolation to systems between the two extreme cases.
\end{abstract}

\maketitle

Since its first experimental observation in 1925\,\cite{PhysRev.26.44}, the HeH$^+$ molecular ion, the simplest {\it polar heteronuclear} molecule, has served as a fundamental benchmark system for understanding principles of molecular formation and electron correlation\,\cite{Banyard1970}. Moreover, HeH$^+$ continues to intrigue researchers as the first molecular species to arise in the universe, but has mysteriously remained absent in astronomical spectra\,\cite{Lepp2002}. Despite its fundamental nature and broad-ranging importance, the behavior of HeH$^+$ in strong fields is largely unexplored, until now.

So far, H$_2^+$, the simplest molecule, has typically been used as the prototype for the dynamics of molecules in strong laser fields\,\cite{Posthumus2004}. This understanding is then extended to interpret and predict the behavior of more and more complex molecules. However, the dynamics of simple molecules such as H$_2^+$\,\cite{Ben-Itzhak2005}, N$_2$\,\cite{Voss2004}, and O$_2$\,\cite{Sayler2007} lacks several fundamental properties as these homonuclear molecules are symmetric and do not have a permanent dipole moment.  In stark contrast, HeH$^+$ is a two-electron system with a large mass asymmetry, a strong electronic asymmetry, and a permanent dipole moment, as depicted in Fig.\,1. Most polar molecules, for example HCl or CO \cite{Akagi2009,Cornaggia1991,Zigo2017}, lie somewhere between the perfect symmetry of H$_2^+$ and extreme asymmetry of HeH$^+$. Therefore, knowledge of both is required for the general understanding of laser-induced molecular dynamics, which is necessary to form a foundation from which the behavior of more complex molecules can be predicted.

The reluctance to address HeH$^+$ in strong fields \cite{FelHeH} is rooted in difficulties on both the experimental and the theoretical side. Experimentally, since neutral HeH is not stable, the ion must be synthesized, e.g. in a plasma, and prepared in an ion beam apparatus. 
Furthermore, the resulting low target density and the large intensities needed to remove an electron from HeH$^+$ ($I$\,$\approx$\,$10^{16}\,{\rm W/cm}^2$) result in extremely low event rates and long measurement times.

On the theory side, the solution of the time-dependent Schr\"odinger equation (TDSE) with all 9 degrees of freedom necessary to describe HeH$^+$ is extremely challenging and unfeasible with current computational resources. Therefore, approximations such as TDSE calculations with reduced dimensionality\,\cite{Dehghanian2013,Li2016,Wang2017} or semiclassical approaches\,\cite{Vila2017} are often used.

\begin{figure}[htbp] \centering
\includegraphics[width=1\columnwidth]{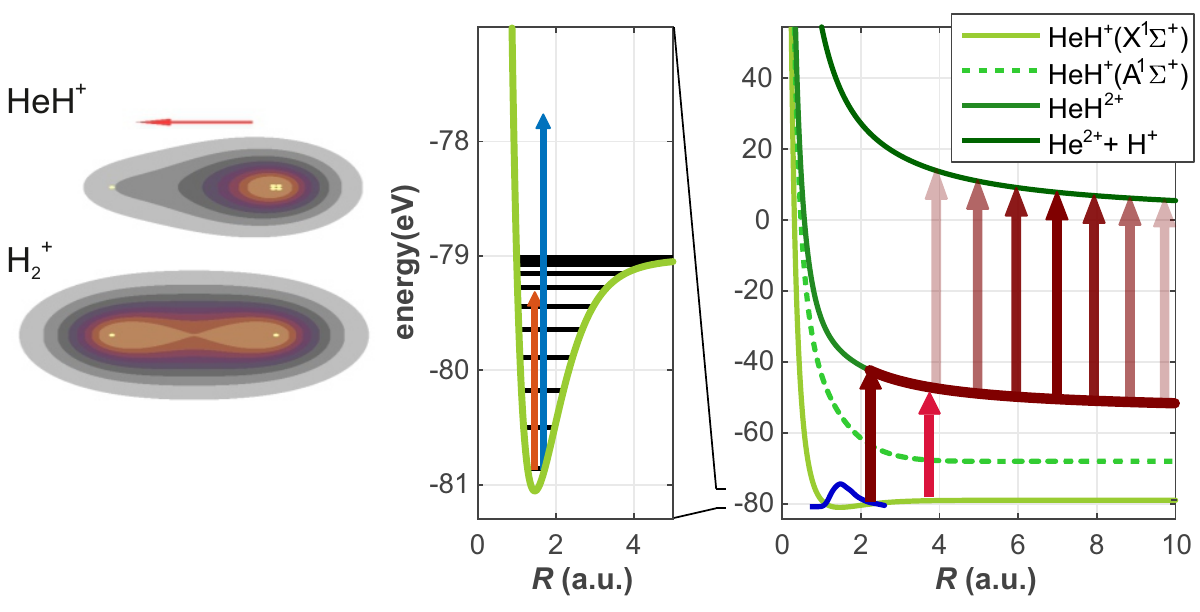}
\caption{Left: Extremal cases of molecular bonds: The highest occupied molecular orbital for the asymmetric HeH$^+$(top), where the direction of the dipole moment is depicted by the red arrow, compared to the situation of the perfectly symmetric molecule H$_2^+$ (bottom). In the right panel, the potential energy curves of HeH$^+(X^1\Sigma^+)$, HeH$^+(A^1\Sigma^+)$, HeH$^{2+}$, and HeH$^{3+}$ are displayed. Also shown is the population density of the ground state (blue curve) and the most likely ionization pathways. At low intensities, HeH$^+$ needs to stretch to 4\,a.u. before ionization sets in (red arrow). Double ionization requires much higher intensities. The first ionization step, therefore, can take place already at slightly more than 2\,a.u. while stretching to 10\,a.u. may occur before the second ionization step (dark red). The middle panel zooms into the ground state and displays its vibrational levels. An 800-nm laser pulse (red arrow) will couple these in contrast to a 400-nm pulse (blue).
}
\label{fig:pathways}
\end{figure}

Here we present measurements of the fragmentation of a beam of HeH$^+$ in ultrashort laser pulses. This includes single and double ionization as well as an estimation of nonionizing dissociation for different laser intensities. The results are interpreted using solutions of the reduced-dimensional TDSE and simulations based on dressed surface hopping (DSH). Moreover, we use another peculiarity of HeH$^+$, namely, the existence of isotopologues with large differences in mass asymmetry, along with two laser wavelengths to elucidate the dynamics and substantiate our interpretation. 

The measurements are based on a 10-keV, transversally cold ion beam and 3D coincidence momentum spectroscopy\,\cite{Ben-Itzhak2005,Rathje2013,Wustelt2015}. The HeH$^+$ ions are produced in a duoplasmatron ion source \cite{Liebl1976} using a mixture of helium and hydrogen gas \cite{Ketterle1988} and accelerated to 10\,keV  kinetic energy. A Wien filter, several Einzellenses and apertures are used to select the $q/m$ ratio of the ionic target and produce a transversely cold and well-collimated ion beam in the interaction region with the laser. The laser system delivers 10-mJ, 35-fs pulses at a wavelength of 800\,nm or 3-mJ, 50-fs pulses at 400\,nm. Both neutral and ionic fragments of the HeH$^+$ ion beam are detected in coincidence using a time- and position-sensitive detector, which allows for the retrieval of the full 3D momenta of all nuclear fragments. Further, a weak static longitudinal electric field in the interaction region is employed to separate different fragments by their time of flight. This provides access to the charge state, kinetic energy release (KER), and alignment of each fragmented molecule. In front of the detector the nonfragmented beam was blocked by a Faraday Cup, which affected the detection of the low-energy  neutral fragments of dissociation. The low target density in this measurement ($\approx 10^6/{\rm cm}^3$) limits the event rate to $\lesssim1$\,Hz and requires measurement times of $\gtrsim$1\,week.

The focus of this work is naturally on the question of how does a strong charge asymmetry and the resulting permanent dipole, like that found in HeH$^+$, affect ionization dynamics in strong laser fields? The large permanent dipole moment allows direct transfer of population from lower to higher vibrational states, which results in stretching of the molecule on the femtosecond timescale\,\cite{Li2016}. Electronic excitation, on the other hand, is negligible\,\cite{Ursrey2012} due to the large energy difference between the electronic ground state $X^1\Sigma^+$ and the first excited state $A^1\Sigma^+$, see Fig.\,1. This situation is precisely the opposite  to that in homonuclear molecules, like H$_2^+$, where the initial stretching of the molecular bond proceeds exclusively via electronic excitation without direct vibrational transitions. Accordingly, nonionizing dissociation, which typically proceeds via electronic transitions, should be extremely low for HeH$^+$. Because of the low yield, low KER, and the exceedingly difficult task of separating neutral He from the ion beam, the dissociation probability can only be estimated here. Nevertheless, the small dissociation to ionization branching ratio, which is $<$ 5\% at 800\,nm and below detection levels at 400 nm, confirms the assertion bolstered below --  vibrational coupling dominates electronic coupling for HeH$^+$.

The intensity- and KER-dependent single ionization yield of HeH$^+$ at 800\,nm pulses is shown in Fig.\,2. Because of the large energy difference between dissociation limits of HeH$^{2+}$, only the charge-symmetric channel, ${\rm HeH}^+ + n\hbar\omega \longrightarrow {\rm He}^+ + {\rm H}^+ +e^-$, is observed. Here, the most distinct feature is a transition region around $2\times 10^{15}\,{\rm W/cm^2}$, highlighted by an arrow. To understand this feature, we solve the TDSE whereby the dynamics of the nuclei and the active electron are restricted to one dimension\,\cite{TDSE}\nocite{Green1974,Feit1982}. The simulations (dashed curves in Fig. 2) are averaged to match the intensity distribution in the laser focus and the population of the lowest vibrational states\,\cite{Loreau2011}. In order to understand the molecular dynamics, it is expedient to convert the KER into the internuclear distance $R$ at the instant of ionization using the Born-Oppenheimer potential curves for HeH$^{2+}$, see upper abscissa axis in Fig.\,2. At the lowest intensities, ionization reaches a maximum at $R \approx 3.7$\,a.u.. However, some of the molecules are stretched up to 6\,a.u. at the instant of ionization. At the highest intensities, in contrast, ionization peaks at $R \approx 2.2$\,a.u. and the largest observed internuclear distance does not exceed 4\,a.u.

The dynamics responsible for these observations are as follows: At low intensity, ionization by 800-nm pulses starts with 1-photon vibrational excitation from the initial population $v = 0-1$ to excited states, $v = 6-10$ in the rising part of the laser pulses, see Fig.\,2(c). This allows the molecule to stretch, which in turn significantly reduces the ionization potential from 41\,eV ($R = 1.75\,{\rm a.u.}$) to 29\,eV ($R = 6\,{\rm a.u.}$), see Fig.\,1. An estimate of the characteristic time of this stretching motion can be done by the vibrational period associated to the $v = 6-10$ states, which is on the order of 34-100\,fs or by the time a corresponding classical particle moves from $R$ = 2 to $R $= 3.7\,a.u., which is $\approx$12\,fs. As a consequence, the ionization rate increases sharply for the stretched molecules\,\cite{Dehghanian2013}, which induces molecular vibration and further ionization via the Lochfra{\ss} mechanism\,\cite{Goll2006}. This mechanism is corroborated by the eye-catching lack of yield at KER$ = 8.5$ \,eV ($R=$ 3.1 a.u.) in the TDSE results and also the measurements, see Fig.\,2. This peculiar feature appears to be a product  of the vibrational wave functions and happenstance.  Namely, the dominant excited vibrational state wave functions have overlapping nodes at this internuclear distance, which results in the aforementioned lack of yield at KER$ = 8.5$\,eV.

At higher intensities, $I>2\times 10^{15}\,{\rm W/cm^2}$, the KER peaks at approximately 11\,eV (indicating $R \approx 2.2$\,a.u.), while the maximum KER is approximately 15\,eV ($R \approx$ 1.75\,a.u.), see Fig.\,2(a). This suggests that the higher intensity is sufficient for direct ionization without vibrational excitation and stretching as required at lower intensities. The measured value is slightly larger than the equilibrium internuclear distance, $R\approx$ 1.5\,a.u., as the outer turning point of the initial vibrational wave packet is preferentially ionized due to the sharp decrease in ionization potential with increasing $R$.
\begin{figure}[htbp] \centering
\includegraphics[width=.95\columnwidth]{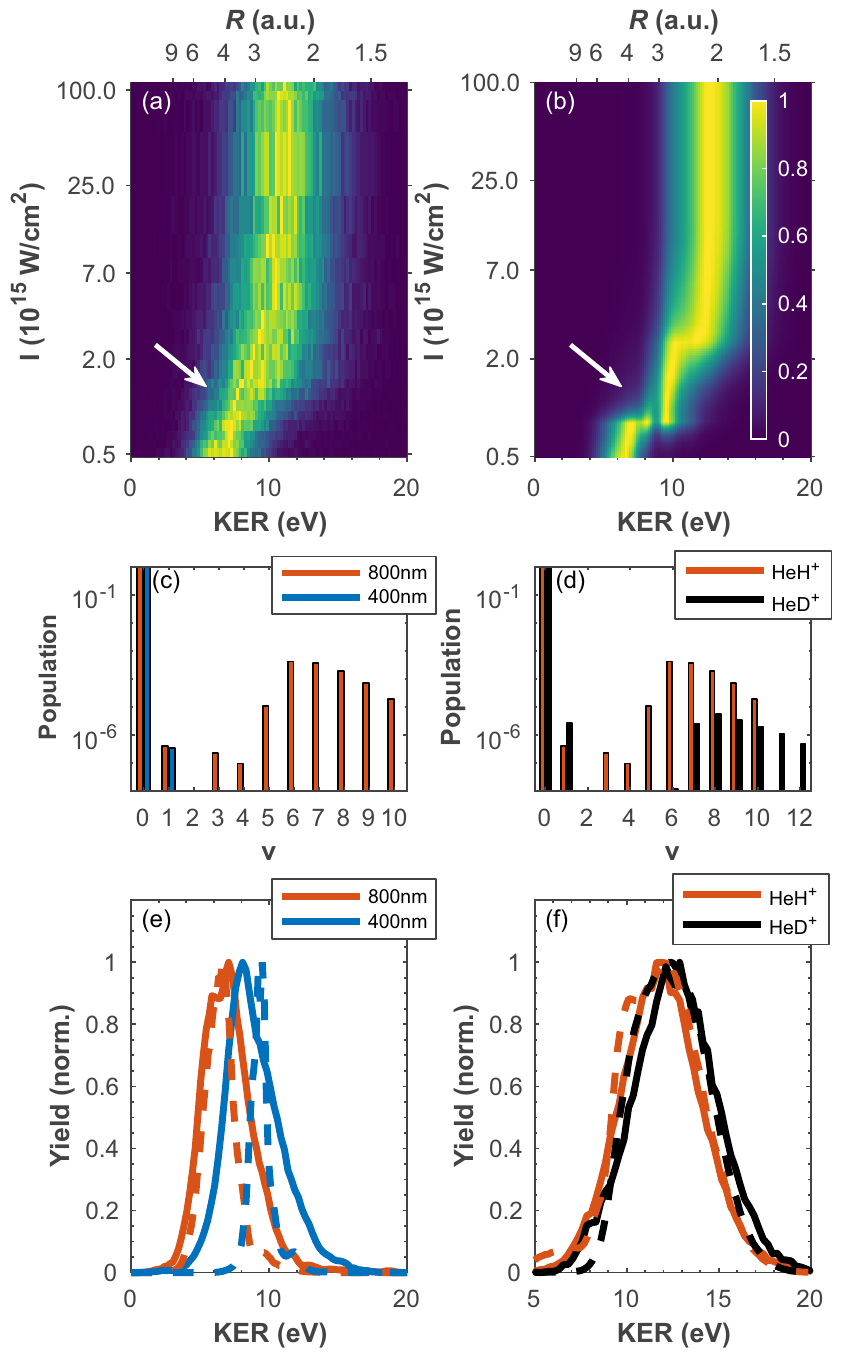}
\caption{Single ionization of HeH$^+$. (a) Measured fragmentation yield as a function of intensity (log. scale) and KER (lower axis) or internuclear distance $R$ (upper axis) for 800-nm pulses. (b) Same as (a) for focal volume and vibrationally averaged TDSE results (I$_{max}$\,=\,10$^{16}$\,W/cm$^2$). The yield is normalized for each intensity. At $I< 10^{15}\,{\rm W/cm^2}$ substantial ionization is only possible after the bond has stretched to $R\gtrsim 4$\,a.u., while HeH$^+$ can be ionized without appreciable stretching for $I \ge 10^{16}\,{\rm W/cm^2}$. At 800\,nm, the higher vibrational states can be resonantly excited. This can be seen in panel (c), where computed population probabilities of the vibrational states of HeH$^+$ are displayed after interaction with 800-nm and 400-nm pulses ($10^{15}\,{\rm W/cm^2}$), which is representative for the excitation dynamics during the pulse where ionization quickly ionizes the population in the higher vibrational states. The correspondingly reduced stretching dynamics at 400\,nm (cf.\,\cite{Li2016} for calculations in the context of high-harmonic generation at 600\,nm) results in slightly larger KER at low intensities. This is shown in panel (e), which compares KER spectra for both wavelengths at $10^{15}\,{\rm W/cm^2}$. The dashed curves pertain to TDSE simulations. Stretching can also be impeded by increasing the reduced mass, which is accomplished by using the HeD$^+$ isotopologue in panel (f), where the KER spectra for 800\,nm are displayed at a peak intensity of $\approx 10^{17}\,{\rm W/cm^2}$. The larger mass of HeD$^+$ inhibits stretching, which results in less population of the excited vibrational states as compared to HeH$^+$, see (d). Note, that for the larger reduced mass of the molecule, higher vibrational levels were preferentially excited and correspond to similar extent along $R$.}
\label{fig:IscanSI}
\end{figure}
To strengthen the assertion that the 1 photon excitation of higher vibrational states starts the initial stretching of the molecule, we have also performed the measurements with 400-nm laser pulses. As seen in Fig.\,1 at this wavelength, the photon energy is too large to excite higher bound vibrational states. Therefore, the number of events with low KER, which requires vibrational excitation, is significantly reduced at low intensity, see Fig.\,2(e), while the KER spectra at high intensity, where direct ionization dominates, are very similar for both wavelengths. Moreover, the TDSE calculations confirm these results showing that the excitation of higher vibrational states is indeed much stronger for 800\,nm, see Fig.\,2(c). In contrast, for larger energies (KER$ \approx $9\,eV), which corresponds to a smaller internuclear distance ($R \approx $3\,a.u.) near the outer turning point of the initial wave function, 400\,nm is predicted to more effectively drive electronic excitation and enhance ionization as compared to 800\,nm \cite{Dehghanian2013,Wang2017a}. Note that \cite{Dehghanian2013} uses fixed-nuclei approximations. In contrast to their work, we highlight the relevance of vibrational excitation, which is the prerequisite step and is not discussed in Ref.\,\cite{Dehghanian2013}.

We are further able to understand and manipulate the dynamics by using $^4$HeD$^+$, which effectively slows down the fragmentation process by increasing the reduced mass, $\mu=m_1 m_2/(m_1 + m_2)$, by 67\%. Unfortunately, in addition to the aforementioned experimental challenges, measurements with HeD$^+$ are even more demanding because the synthesis of HeD$^+$ inevitably produces a large fraction of D$_3^+$, which has virtually the same molecular mass. A pure $^4$HeD$^+$ ion beam can therefore not be realized. Nevertheless, by taking advantage of coincidence detection and the high precision of our apparatus, we were able to discriminate D$_3^+$ fragments by applying an electrostatic field in the interaction region and relying on the extremely small, 6:1000, mass defect. In addition to its larger reduced mass, HeD$^+$ has a smaller dipole moment compared to HeH$^+$. Both aspects slow down stretching, as can be seen in the reduced population of higher vibrational states for HeD$^+$, see Fig.\,2(d), with the consequence that higher intensities are required for ionization as the ionization potential increases with decreasing internuclear distance. In agreement with the proposed mechanism of ionization of helium hydride, we find higher KERs in HeD$^+$ as compared to HeH$^+$, see Fig.\,2(f).

 \begin{figure}[htbp] \centering
\includegraphics[width=1\columnwidth]{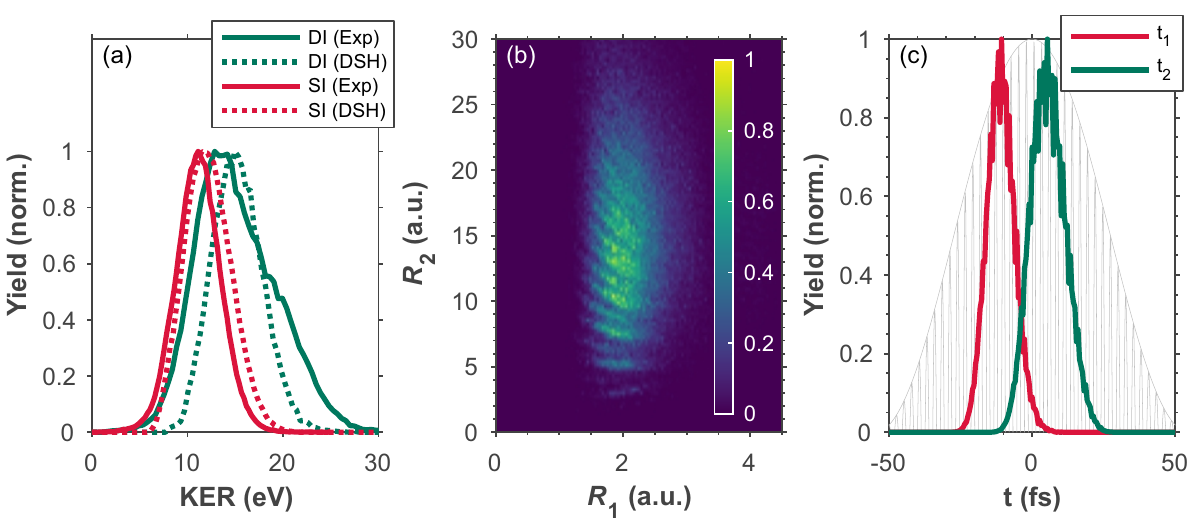}
\caption{Double ionization of HeH$^+$: (a) Measured KER-dependent double ionization yield (solid green line) and, for comparison, the single ionization yield (solid red line) at I$\approx 10^{17}\,{\rm W/cm^2}$ peak intensity. The corresponding computed spectra are represented by dashed curves (I$_{max}$\,=\,6$\cdot$10$^{15}$\,W/cm$^2$). (b) Joint $R_1$-$R_2$ distribution extracted from DSH simulations. The first ionization step occurs at $R_1 \approx 2$\,a.u., while the second electron can be ionized during an extended time interval as reflected by the broad $R_2$ distribution, see also Fig. 1. The fringes reflect the fact that the second ionization takes place only at the extrema of the laser field in each half-cycle. Panel (c) displays double ionization of HeH$^+$ in the time domain. The gray curve represents the instantaneous laser intensity. First and second ionization are plotted in the same colors as in panel (a).}
\label{fig:DI}
\end{figure}

At an intensity above $10^{16}\,{\rm W/cm^2}$, double ionization of HeH$^+$ can also be observed. Since it is currently not feasible to solve the TDSE for a two-electron problem at such high intensity, we resort to semiclassical dressed surface hopping (DSH) simulations in a sequential ionization process. The nuclei are treated as classical particles moving on the instantaneous ground state field-dressed potential surfaces of HeH$^+$ and its daughter ions, with inclusion of their permanent dipole moments and the molecular polarizabilities. Jumps to singly- and doubly-ionized surfaces are determined by the $R$- and orientation-dependent ionization rates implemented with many-electron weak-field-asymptotic theory (ME-WFAT)\,\cite{Tolstikhin2014} with an active-space partition scheme\,\cite{Yue2017}. The initial classical phase space of the random trajectories were chosen according to a Husimi distribution. In addition to being computationally much less demanding, the DSH approach enables tracing of molecular dynamics leading to double ionization by analyzing the pathway\textbf{s} of the individual trajectories, which yields insights into the single ionization and double ionization process not available from the TDSE. Electron rescattering and vibrational couplings are not accounted for in the model. Details of the simulation technique will be presented in an upcoming publication\,\cite{Lun2018}.

At the intensities in question, good agreement of measured and computed KER spectra is achieved for single and double ionization; see Fig.\,3(a). Noteworthy is the broad width of the KER spectra for double ionization, indicating an extended time window for the second ionization step. The analysis of the theoretical data allows more detailed conclusions to be drawn about the dynamics of double ionization. In Fig.\,3(b) the joint distribution of the internuclear distance at the first and second ionization step, $R_1$ and $R_2$, is displayed. As already discussed, the first electron is ionized at $R_1 \approx 2$\,a.u. at high intensity. The second electron can only be released at subsequent field extrema, which manifests itself in the stripelike $R_1$-$R_2$ distribution. The second ionization yield reaches its maximum at $R_2\approx 10$\,a.u. more than 10\,fs after the first ionization, see also Fig.\,3(c) and Fig.\,1(a). Not unlike the single-ionization scenario, the reason for this process is that the ionization potential decreases with increasing internuclear distance while the field strength of the laser rises.

In conclusion, by exploring the laser-induced fragmentation of helium hydride we are establishing the heteronuclear limit of strong-field molecular dynamics. The extreme asymmetric nature of HeH$^+$, which is manifested in a strong permanent dipole, results in fundamentally different fragmentation dynamics than those seen in homonuclear molecules. Namely, direct vibrational excitation, with almost no electronic excitation as the initial process, dictates the fragmentation process here, which is the antithesis of symmetric molecules. Together with the understanding of simple symmetric molecules like H$_2^+$, our findings bookend the spectrum of molecular symmetry, giving a valuable new insight in heteronuclear systems.

The authors acknowledge funding by the Deutsche Forschungsgemeinschaft (DFG) in the frame of the Schwerpunktprogramm (SPP) 1840, Quantum Dynamics in Tailored Intense Fields. 	

\bibliography{natbib_arxiv}

\pagebreak

\onecolumngrid
\begin{center}
  \textbf{\large Supplementary materials for:\\ Heteronuclear Limit of Strong-Field Ionization: Fragmentation of HeH$^+$ by Intense Ultrashort Laser Pulses}\\[.2cm]

  P. Wustelt,$^{1,2}$ F. Oppermann$^{3}$, Lun Yue$^{4}$, M. M\"oller,$^{1,2}$,\\
 T. St\"ohlker$^{1,2}$, 
S. Gr\"afe$^{4}$, M. Lein$^{3}$, G. G. Paulus$^{1,2}$, A. M. Sayler$^{1,2}$\\
{\itshape$^{1}$ Institute of Optics and Quantum Electronics, Friedrich Schiller University Jena, 07743 Jena, Germany}\\
{\itshape$^{2}$ Helmholtz Institute Jena, Fr\"obelstieg 3, 07743 Jena, Germany}\\
{\itshape$^{3}$ Institut f\"ur Theoretische Physik, Leibniz Universit\"at  Hannover, Appelstra{\ss}e 2, 30167 Hannover, Germany}\\
{\itshape$^{4}$ Institute of Physical Chemistry, Friedrich Schiller University Jena, Helmholtzweg 4, 07743 Jena, Germany}\\
(Dated: \today)\\[1cm]
\end{center}
\twocolumngrid

\setcounter{equation}{0}
\renewcommand{\theequation}{S\arabic{equation}}

\section{Time-dependent Schr\"odinger equation}

The time-dependent Schr\"odinger equation (TDSE) is solved for a one-dimensional single active electron model of HeH$^+$. In order to overcome the Born-Oppenheimer (BO) approximation, this model consists of coupled one-dimensional motion of the nuclei and one-dimensional motion of the electron. Rotation of the molecule is not considered, the electric field polarization is assumed to be along the molecular axis. Experimental results show that most dynamics happen when the molecule is aligned to the laser polarization direction. Only single ionization can be treated.

The TDSE reads
\begin{equation}
    i \hbar \frac{\partial \psi(x,R,t)}{\partial t} = \left(H_0 + (\kappa x + \lambda R)eE(t) \right) \psi(x,R,t),
\end{equation}
where \(H_0\) is the field-free Hamiltonian
\begin{equation}\label{eq:field_free_Hamiltonian}
    \begin{aligned}
        H_0 = \frac{p^2}{2 \mu_e} + \frac{P^2}{2 \mu} & - \frac{Z_1 e^2}{\sqrt{(x + \frac{m_2}{m_n}R)^2 + \alpha_1(R)}} \\
             + V_{\text{ion}} (R) & - \frac{Z_2 e^2}{\sqrt{(x - \frac{m_1}{m_n}R)^2 + \alpha_2(R)}}.
    \end{aligned}
\end{equation}
and \(\kappa = (m_n + Z_1 + Z_2)/(m_n + 1)\), \(\lambda = (Z_1 m_2 - Z_2 m_1)/m_n\). Here \(x,p\) and \(R,P\) are coordinate and momentum of the active electron and the internuclear separation respectively, \(m_i\) and \(Z_i\) are the nuclear masses and the effective charges on the helium and proton side respectively, \(m_n = m_1 + m_2\). The inactive electron is assumed to be located at the helium core, i.e. $Z_1 = Z_2 = 1$. \(V_\text{ion}(R)\) is the potential energy when the active electron is removed, i.\,e. the ground state BO potential energy curve of HeH\({}^{2+}\). 

The parameters for the softened Coulomb potential, \(\alpha_i(R)\), are tuned for each internuclear distance such that the two lowest BO curves are reproduced exactly\,\cite{Green1974}. This basically makes the potential well deeper at the helium than at the hydrogen.
For the time evolution, the initial state is chosen to be one of the
low-lying eigenstates of \(H_0\) corresponding to vibrational states in
the electronic ground state of the molecule. It is then propagated on a grid using the split operator method \cite{Feit1982} with Fourier transforms.
When the wave function approaches the boundary of the grid in \(x\) direction, it is absorbed and from then on propagated without electron-nuclei interaction.
In most of the calculations, the absorbing boundary starts at 179\,a.u. distance from the nuclear center of mass.

The probability distribution for the KER after single ionization is calculated from the absorbed part of the wave function at the end of the time evolution.

\end{document}